\documentclass{aa}  
\usepackage{graphicx}
\usepackage{txfonts}
\usepackage{natbib}
\usepackage{hyperref}

\usepackage{color}

\graphicspath{{./}{figures/}}

\begin{document}

\title{Source positions of an interplanetary type III radio burst and anisotropic radio-wave scattering}


 \author{Xingyao Chen\inst{\ref{inst1}}, 
 Eduard P. Kontar\inst{\ref{inst1}} , 
 Nicolina Chrysaphi\inst{\ref{inst2},\ref{inst1},\ref{inst3}}, 
 Peijin Zhang\inst{\ref{inst4}}, 
 Vratislav Krupar\inst{\ref{inst5},\ref{inst6}}, 
 Sophie Musset\inst{\ref{inst7}}, 
 Milan Maksimovic\inst{\ref{inst3}}, 
 Natasha L. S. Jeffrey\inst{\ref{inst8}}, 
 Francesco Azzollini\inst{\ref{inst1}},
 Antonio Vecchio\inst{\ref{inst3}}
 }

\institute{
School of Physics \& Astronomy, University of Glasgow, Glasgow, G12 8QQ, UK \label{inst1}
\email{Xingyao.Chen@glasgow.ac.uk}
\and
Sorbonne Universit\'{e}, \'{E}cole Polytechnique, Institut Polytechnique de Paris, CNRS, Laboratoire de Physique des Plasmas (LPP), 4 Place Jussieu, 75005 Paris, France \label{inst2}
\and
LESIA, Observatoire de Paris, Universit\'{e} PSL, CNRS, Sorbonne Universit\'{e}, Universit\'{e} de Paris, 5 place Jules Janssen, 92195 Meudon, France \label{inst3}
\and
Department of Physics, University of Helsinki, P.O. Box 64, Helsinki, Finland\label{inst4}
\and
Goddard Planetary Heliophysics Institute, University of Maryland, Baltimore County, Baltimore, MD, USA\label{inst5}
\and
Heliophysics Science Division, NASA Goddard Space Flight Center, Greenbelt, MD, USA\label{inst6}
\and
European Space Agency (ESA), European Space Research and Technology Centre (ESTEC), Keplerlaan 1, 2201 AZ Noordwijk, The Netherlands \label{inst7}
\and
Department of Mathematics, Physics \& Electrical Engineering, Northumbria University, Newcastle upon Tyne, NE1 8ST, UK \label{inst8}
}


\abstract
{Interplanetary solar radio type III bursts provide the means to remotely study and track energetic electrons propagating in the interplanetary medium. Due to the lack of direct radio source imaging, several methods have been developed to determine the source positions from space-based observations. Moreover, none of the methods consider the propagation effects of anisotropic radio-wave scattering, which would strongly distort the trajectory of radio waves, delay their arrival times, and affect their apparent characteristics. We investigate the source positions and directivity of an interplanetary type III burst simultaneously observed by Parker Solar Probe, Solar Orbiter, STEREO, and Wind and we compare the results of applying the intensity fit and timing methods with ray-tracing simulations of radio-wave propagation with anisotropic density fluctuations.
The simulation calculates the trajectories of the rays, their time profiles at different viewing sites, and the apparent characteristics for various density fluctuation parameters. The results indicate that the observed source positions are displaced away from the locations where emission is produced, and their deduced radial distances are larger than expected from density models. This suggests that the apparent position is affected by anisotropic radio-wave scattering, which leads to an apparent position at a larger heliocentric distance from the Sun.
The methods to determine the source positions may underestimate the apparent positions if they do not consider the path of radio-wave propagation and incomplete scattering at a viewing site close to the intrinsic source position.
}


\titlerunning{Source positions of an interplanetary type III burst}
\authorrunning{X.Chen \& E.Kontar et al.}

\maketitle

\section{Introduction}
\label{sec-intro}

Radio bursts at kilometer and hectometer wavelengths are generated by energetic electrons propagating in interplanetary space through the plasma emission mechanism \citep{1958SvA.....2..653G, 1985srph.book.....M, 1985ARA&A..23..169D, 1987SoPh..111...89M}. This mechanism produces fundamental and second harmonic emissions at the local plasma frequencies $f_\mathrm{pe}$ and $2f_\mathrm{pe}$, respectively, where $f_\mathrm{pe}=\sqrt{e^2 n (r)/\pi m_e}$ is determined by the electron plasma frequency, electron number density, electron charge, and mass. The theoretical radial heliocentric distance can be calculated based on the emission frequency and assuming an interplanetary density model.

Interplanetary (IP) type III bursts provide valuable information on the electron beam trajectory, density distribution, and magnetic field configuration from the solar corona to the interplanetary medium. However, direct imaging of radio emission in the frequency range of radio waves affected by the ionosphere (below about 20 MHz) is challenging, and interferometric imaging from multiple spacecraft has yet to be available. Various methods have been developed to determine the source positions of these bursts.

Direction-finding (DF) capabilities -- also referred to as GonioPolarimetric (GP) capabilities -- of radio receivers carried on spacecraft can retrieve the direction, polarization, and flux of incoming electromagnetic radio waves \citep{1972Sci...178..743F, 1998JGR...10329651R, 2005RaSc...40.3003C, 2008SSRv..136..549C, 2014SoPh..289.4633K}. Two types of DF techniques exist: spinning demodulation GP, developed for spinning spacecraft observations such as ISEE-3 and WIND/WAVES \citep{1995SSRv...71..231B}, and instantaneous GP, developed for three-axis stabilized spacecraft such as Cassini/RPWS \citep{2004SSRv..114..395G} and STEREO/Waves \citep{2008SSRv..136..487B} \citep{1972Sci...178..743F, 1974SSRv...16..145F, 1980SSI.....5..161M,1985A&A...153..145F, 1995RaSc...30.1699L, 2005RaSc...40.3003C, 2008SSRv..136..549C, 2012SoPh..279..153M, 2012JGRA..117.6101K}. 
These antennas measure the electric field of passing electromagnetic waves, and from the DF analysis, the directions, wave flux, polarization (the four Stokes parameters), and source size of the arrival of radio waves can be determined.

For a single spacecraft observation, the direction (e.g., longitude and latitude) of radio waves can be determined from DF analysis, and thus the source positions can be obtained with the use of both the above directions and an interplanetary density model. For more than two spacecraft observations, the radio source is located at the intersection of the line-of-sight directions from the DF analysis of each spacecraft.
It is worth noting that the accuracy of the DF results depends on the spacecraft separation as well as the calibration accuracy of the antenna parameters, including effective electrical lengths, gains, and effective electrical vectors, which is different from assessing how accurately these electrical antenna parameters have been determined from various modeling efforts \citep{2005AdSpR..36.1530R, 2008SSRv..136..529B, 2009SoPh..259..255R}. Currently, only the DF data for STEREO can be publicly accessed. Moreover, the DF technique assumes the free propagation of radio waves and does not consider any propagation effects that may significantly affect the measured positions \citep{2018ApJ...868...79C, 2019ApJ...884..122K}.

In addition to the DF analysis, the time of arrival (ToA) difference is used to derive the trajectory of radio sources, where the centroids of the sources can be determined from two time-delay measurements from three spacecraft \citep{1977SoPh...54..431W,1984A&A...140...39S,2009SoPh..259..255R,2010ApJ...720.1395T}. Hyperbolic curves are then generated by applying relative time delays from each pair of spacecraft data to derive the source locations where the curves intersect. 
It is important to highlight that the DF analysis can determine the trajectory of the source in three-dimensional positions in the interplanetary space, whereas the ToA analysis mostly indicates the projected source position in the ecliptic plane.
Another method to determine the source direction is to fit the peak intensity of the radio waves at four viewing sites, as done by \citet{2021A&A...656A..34M}.
In some cases, when Langmuir waves are observed alongside interplanetary radio bursts, the radio source can be assumed to be close to the spacecraft.
In some previous studies, such as \citet{1984A&A...141...17B}, the DF method was applied to determine the emission directions of interplanetary type III storms from ISEE-3 observations, and their positions were deduced assuming the radio source region rotates rigidly with the Sun. Additionally, the source position can also be deduced by assuming that the radio sources are located along the Parker spiral magnetic field line originating from the associated active region or flaring sites.

 
The first trajectory measurement of an interplanetary type III burst was conducted by \citet{1972Sci...178..743F} using DF methods, where the radio waves were used to trace nonthermal electrons in the interplanetary medium. Previous studies, such as \citet{1998JGR...103.1923R} and \citet{2014SoPh..289.4633K}, used triangulation based on DF measurements and reconstructed the three-dimensional trajectory of type III bursts, which have suggested that the propagation path of electron beams is roughly along the Parker spiral magnetic field lines.


In spacecraft measurements, it is common for the heliocentric distances of interplanetary radio bursts to be larger than the distance suggested by the density model. In a study by \citet{1984A&A...140...39S}, the heliocentric distances of IP type III bursts in a frequency range of 30 to 1980 kHz were found to be considerable, around 2 to 5 times the local plasma frequency. The authors suggested two possible explanations: scattering or having sources localized in overdense regions. However, their observations seemed to preclude the latter explanation. Another study by \citet{2008SSRv..136..549C} proposed that the larger heliocentric distances could be due to observing the second harmonic component, which radiates at 2$f_{pe}$, a mix of the F and H components, or strong scattering during propagation distorting the path and resulting in a longer heliocentric distance. 
The DF results of a type II burst observed by WIND/WAVES indicated that the azimuth did not intersect the isofrequency contour, suggesting scattering of type II radiation in the interplanetary medium \citep{1998JGR...10329651R}.


The inhomogeneous turbulent solar corona can affect the radio source \citep{2001SoPh..202..131K, 2021NatAs...5..796R} and radio wave propagation \citep[e.g.,][]{2017NatCo...8.1515K,2018ApJ...868...79C, 2019ApJ...884..122K, 2020ApJ...905...43C, 2020ApJ...898...94K, 2021A&A...656A..34M, 2023MNRAS.520.3117C,2023ApJ...946...33C}. In this study, we exclusively concentrate on the impact of this inhomogeneity on radio wave propagation, which may strongly affect solar radio burst properties.
The propagation effects change the direction and path length, as well as cause a delay in the arrival times of radio waves. To understand how local density fluctuations affect the source position, we applied a ray-tracing method to simulate radio-wave propagation for anisotropic density perturbations, taking into account the effects of the Parker spiral model of the interplanetary magnetic field.

Due to the heavy reliance of DF analysis on complex antenna calibrations, which may produce similar results as the timing method, we applied the timing method in a relatively more straightforward way to determine the source positions from both observations and simulations.
In previous studies, \citet{1977SoPh...54..431W} demonstrated that the burst locations determined by DF and time differences were in good agreement. Additionally, \citet{1998JGR...103.1923R} found that the burst profiles measured at Wind and Ulysses closely coincided after making light travel time corrections from the source to spacecraft using the known source locations from the triangulation of the DF analysis of Wind and Ulysses. 
Moreover, \citet{2012ApJ...748...66M} showed that the time difference between the radio wave propagating from source locations (deduced from the triangulation by the DF analysis) to the three spacecraft is consistent with the time-shift between radio flux profiles at the three spacecraft (time-of-flight analysis).

For this study, we first applied the intensity fit and timing method to determine the source positions of an interplanetary type III burst, which has been observed by the radio instruments on board four spacecraft: the Radio Frequency Spectrometer (RFS) on Parker Solar Probe (PSP) \citep{2016SSRv..204...49B, 2017JGRA..122.2836P}, the Radio and Plasma Waves (RPW) instrument on Solar Orbiter (SolO) \citep{2020A&A...642A...1M, 2020A&A...642A..12M, 2021A&A...656A..41M}, WAVES on Solar Terrestrial Relations Observatory (STEREO) \citep{2008SSRv..136..487B}, and the Radio and Plasma Wave Experiment (WAVES) on the WIND spacecraft \citep{1995SSRv...71..231B}.
The radio emission directivity was measured from the intensity distribution at four viewing sites. 
Next, we derived the flux intensity profiles at different viewing angles corresponding to the four spacecraft from the radio-wave propagation simulations with anisotropic scattering effects. The intensity fit and timing method were also applied to determine the source positions from the simulated intensity profiles. We quantitatively investigated and compared the directivity and positions from observations of an interplanetary type III burst and from radio-wave propagation simulations. 

Section \ref{sec-obs} introduces the intensity fit and timing method and presents the source positions deduced from an IP type III burst that was simultaneously observed by four spacecraft. Section \ref{sec-sim} provides the simulation results, including the time profiles at four viewing sites, the apparent source positions, and the source positions deduced from the intensity fit and timing method. Finally, Section \ref{sec-sum} discusses and summarizes the main findings.

\begin{figure*}
\includegraphics[width=9cm]{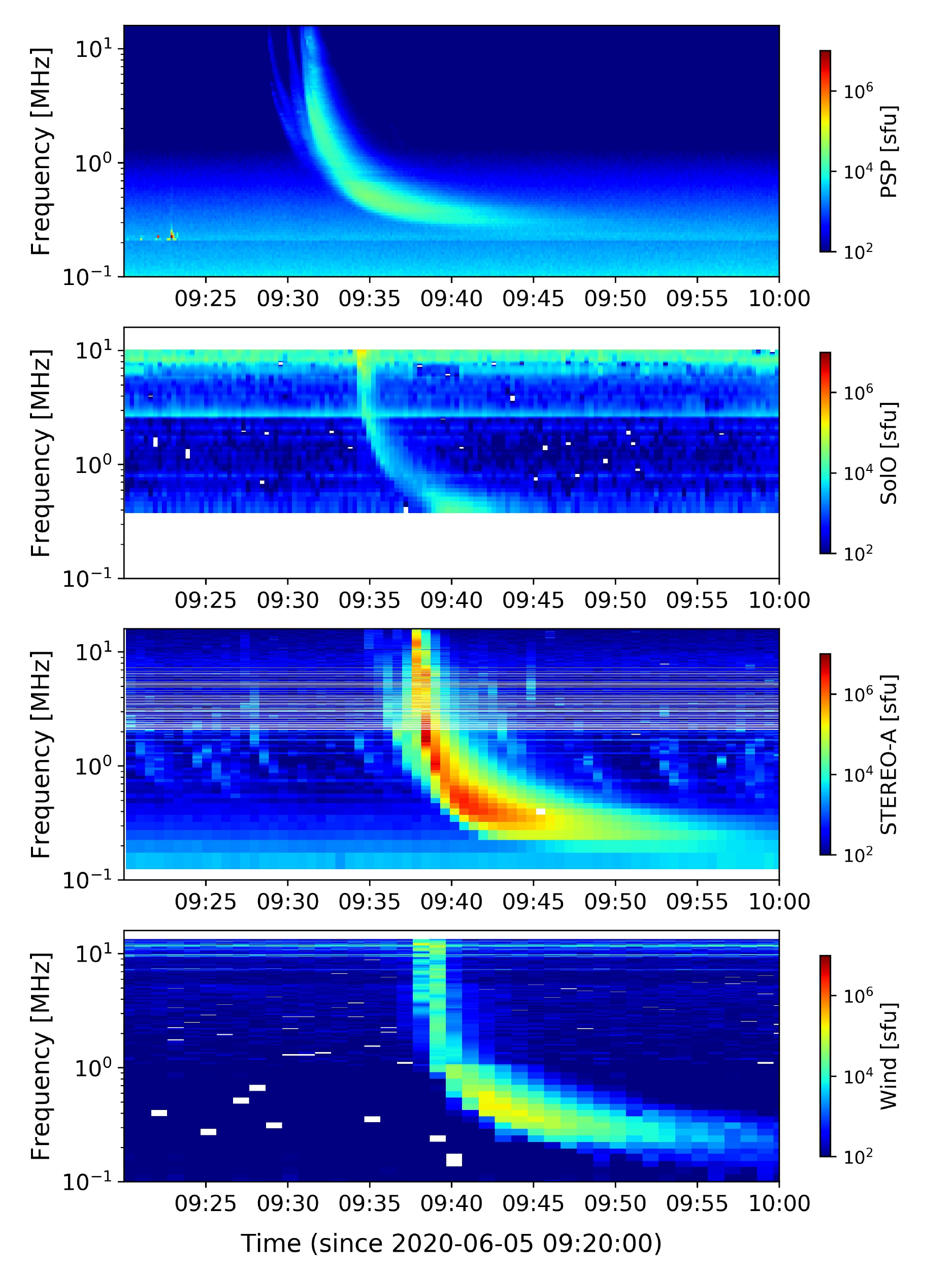}
\hspace{0.1cm}
\includegraphics[width=9cm]{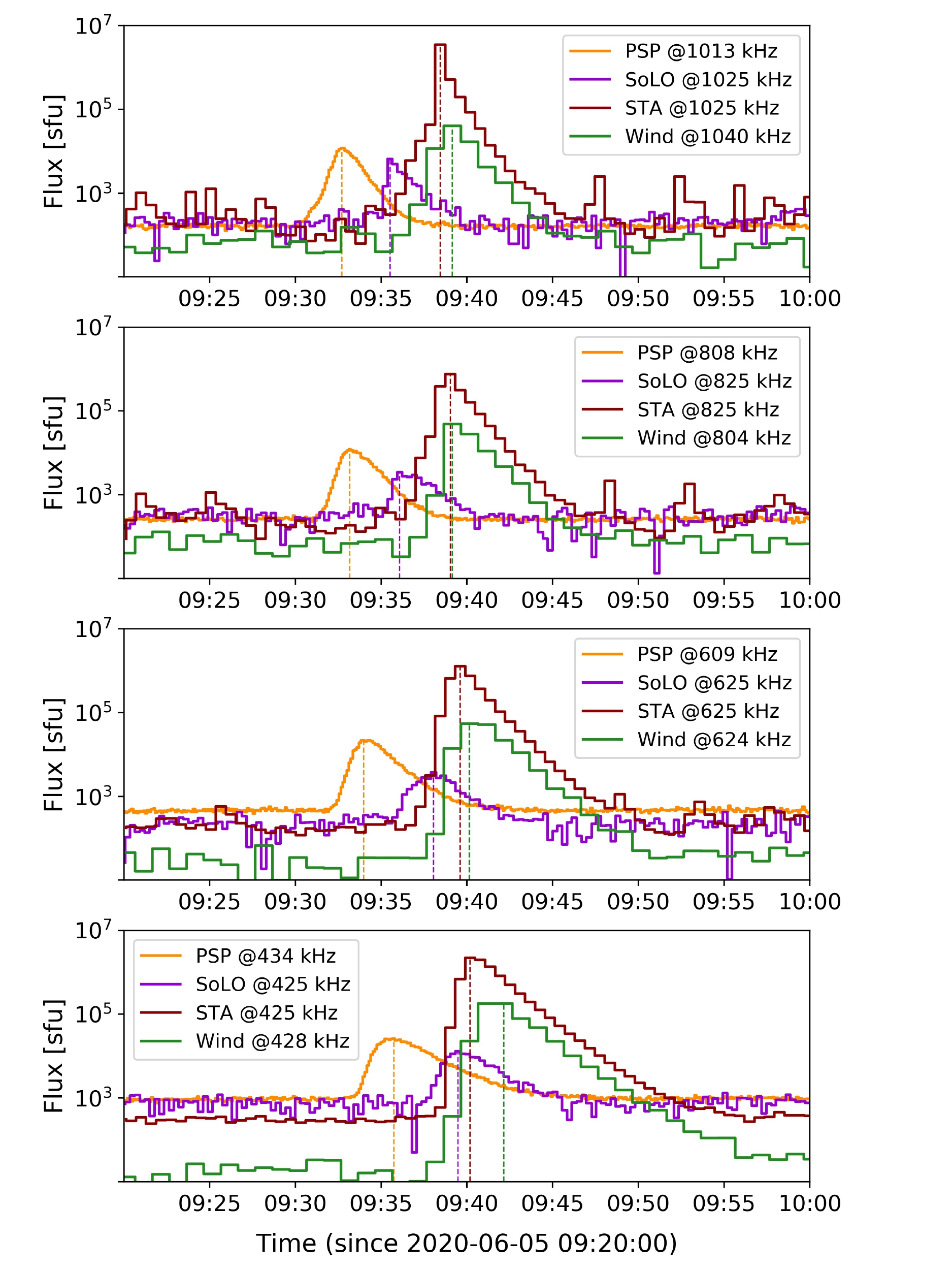}
\caption{Dynamic spectra (left panels) and flux intensity profiles (right panels) of the IP type III burst on 05 June 2020, observed by four spacecraft: PSP, SolO, STEREO-A, and Wind. The dashed vertical lines in the right panels indicate the peak of the time profiles.
\label{obs_spec}}
\end{figure*}

\begin{figure}[h!] 
\centering
\includegraphics[width=9.5cm]{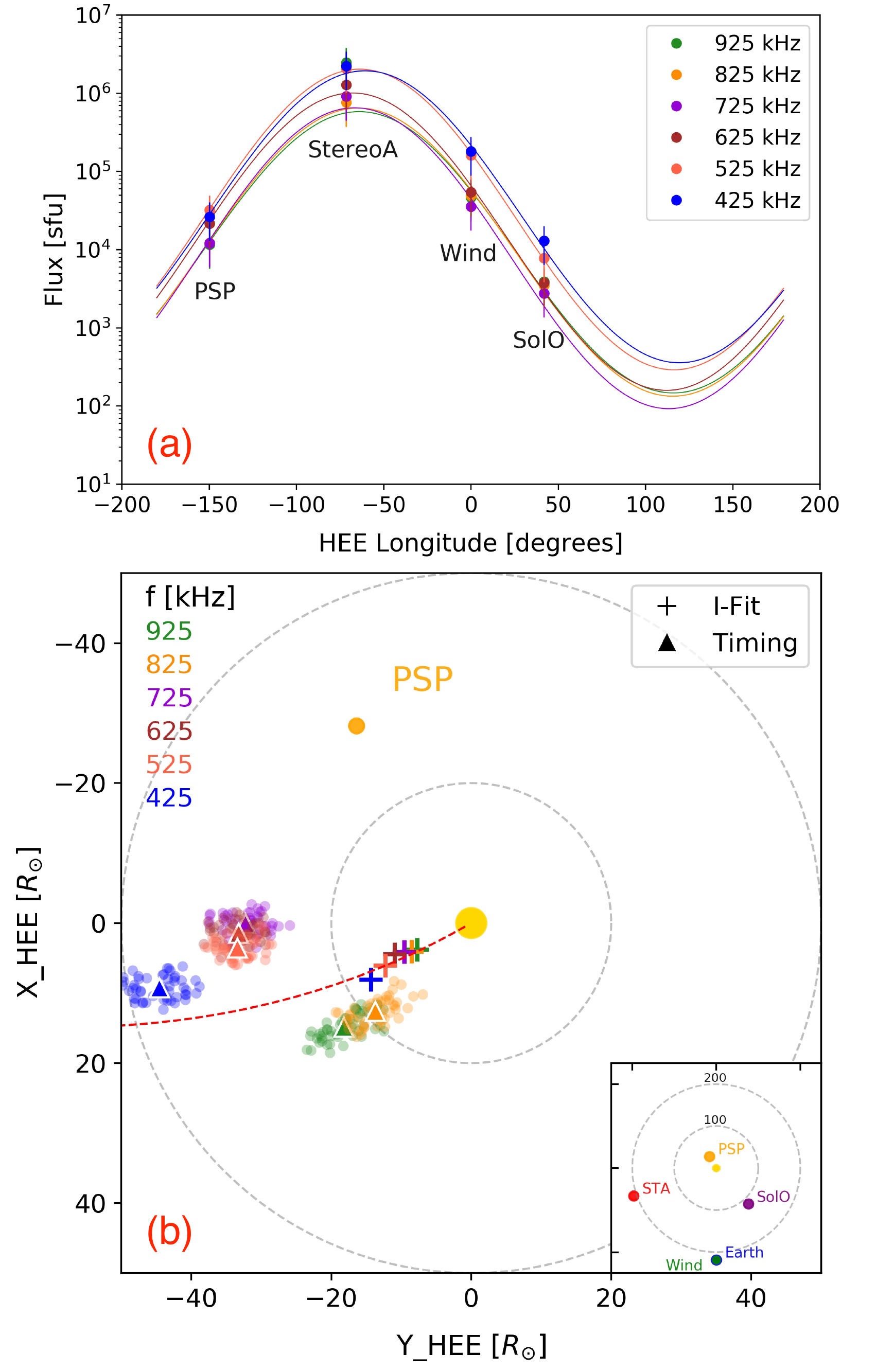}
\caption{Observation results.
(a) Intensity fit result: The peak fluxes at each frequency, corresponding to the longitudes of the four probes, are denoted by points. Vertical lines show the uncertainty, which is given to be 50\% of the peak fluxes. The curves represent the best-fitted fluxes obtained using Equation \ref{eq:I}.
(b) Source positions: The source positions were determined using both the intensity fit (plus symbols) and timing method (triangle symbols) at six frequencies. The shaded dots indicate the positions determined from the timing method by sampling peak times varied by $t^{pk}\pm \Delta t,$ while the triangle symbols show average positions of those sampling dots.
The Parker spiral, with a solar wind speed of 400 km/s, is connected back to the Sun at -60 degrees and indicated as a red dashed line.
In the lower right corner panel, the positions of PSP (P1), SolO (P2), STEREO-A (P3), and Wind (P4) are projected onto the plane of the Earth's orbit in the Heliocentric Earth ecliptic (HEE) coordinate system. The dashed circles represent 100 and 200 times the solar radius.
\label{obs_flux_pos}}
\end{figure}

\begin{figure*}[h!]
\centering
\includegraphics[width=18cm]{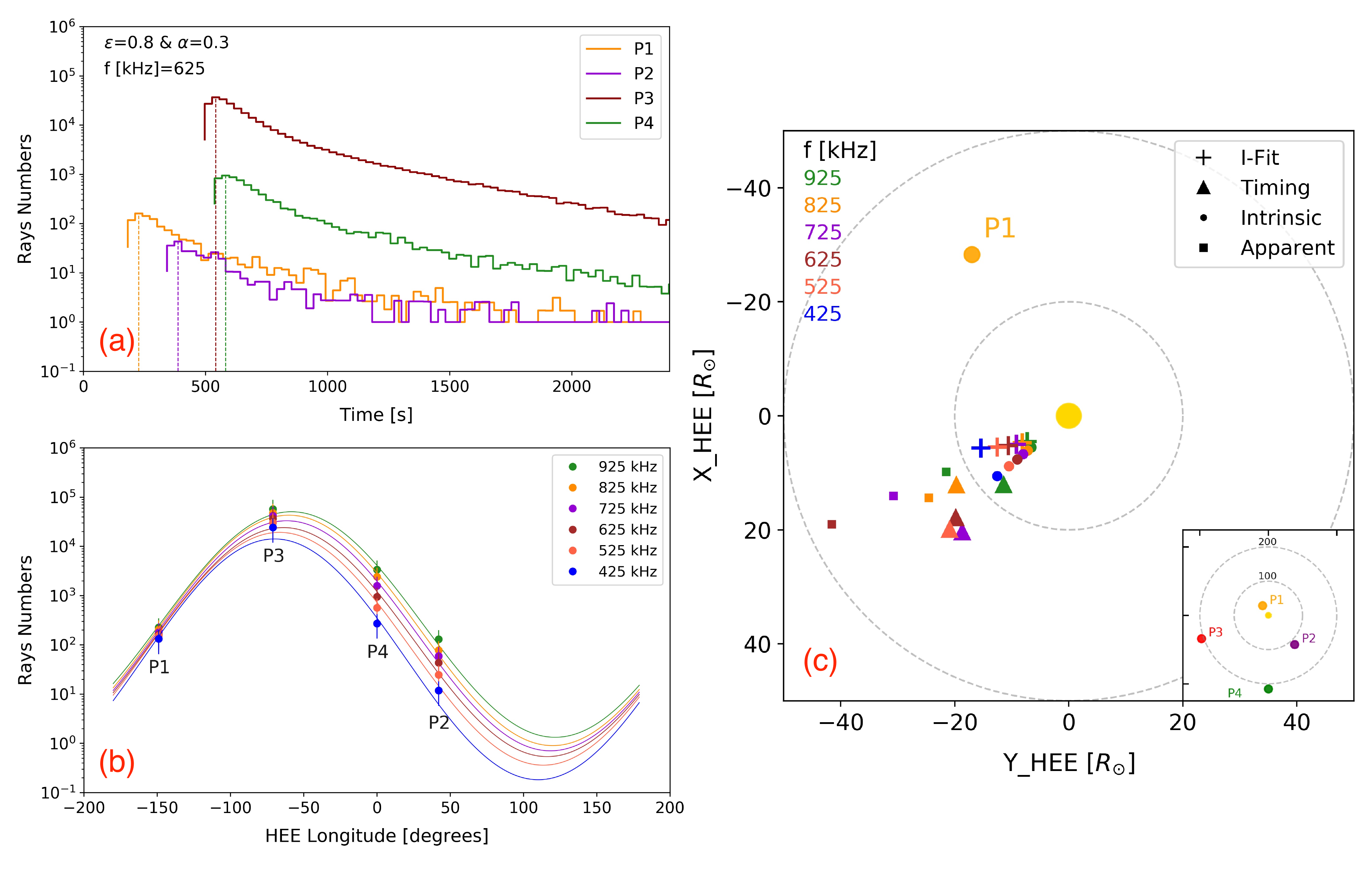}
\caption{Simulation results.
(a) Time profiles at 625 kHz for multiple viewing angles. The collection of rays is restricted to latitudes within the range of 0.85<$cos\phi$<1. The times of the peak flux are marked as vertical dashed lines.
(b) Intensity fit results. The four viewing longitudes are -149$^{\circ}$ (P1), 42$^{\circ}$ (P2), -71$^{\circ}$ (P3), and 0$^{\circ}$ (P4) degrees. The flux errors were set at 50\% of the peak fluxes and are displayed as vertical lines.
(c) Positions of the intrinsic source (initial positions determined based on the density model; solid circle symbols), the apparent source (deduced from scattering simulations; solid square symbols), as well as source positions derived from intensity fits (plus symbols) and timing method (solid triangle symbols).
\label{rays_flux_pos}}
\end{figure*}

\section{Observations}
\label{sec-obs}

The interplanetary radio type III burst was observed around 09:40 UT on 05 June 2020 by four spacecraft. Time resolutions for the observations are approximately 7, 17, 35, and 60 seconds for PSP/RFS, SolO/RPW, STEREO-A/WAVES, and Wind/WAVES, respectively. The positions of the four probes were projected in the plane of Earth's orbit in the heliocentric Earth ecliptic (HEE) coordinate system, as shown in the minor panel of Figure \ref{obs_flux_pos}(b).

The solar radio flux is expected to fall off with distance, approximately $1/R^2$ as the radio waves propagate away from the source. The intensities of the dynamic spectra have been corrected to 1~AU values, while no corrections have been made for the different travel times of the radio waves from the source to the observers. The calibrated dynamic spectra normalized at 1 AU from $\sim$ 200 kHz up to 16 MHz for all four instruments are shown in Figure \ref{obs_spec} (a). We conducted the analysis between 425 kHz and 1025 kHz, as these frequencies are reliable and have been observed well by all four probes. 
The flux curves for the four probes at frequencies close to 425 kHz, 625 kHz, 825 kHz, and 1025 kHz can be seen in Figure \ref{obs_spec} (b). The time profiles follow a quick rise and slow decay, and the rise and decay times increase with decreasing frequency, which is consistent with previous observations of type III bursts. The time bins mark the time resolutions, and the times at peak fluxes are indicated by the vertical dashed lines.

We applied the intensity fit and timing methods to determine the source position of the type III burst propagating through the interplanetary medium.
The PSP spectrum shows multiple type III bursts, comprising an intense type III burst (peaking at $\sim$09:32.5 UT at 1013 kHz) and three weaker type III bursts. Since it can be seen from flux curves that the fluxes of the three weak type III bursts are nearly indistinguishable from the background intensity, which is approximately 100 times lower than the peak fluxes of the intense type III burst, these weak type III bursts had no significant impact on the peak intensity and peak time of the intense type III burst, at least within the provided uncertainty range for the intensity fit and timing methods. For our analysis, we focus on the timing method using the peak time and the intensity fit method using the peak intensity for the intense type III burst.

\subsection{Source positions from radio emission directivity}
\label{sec-obs-Ifit}

The peak flux of the burst at each viewing site can be described by the directivity equation for radio emissions \citep{2008A&A...489..419B, 2019ApJ...873...33B, 2021A&A...656A..34M}:
\begin{equation}
\label{eq:I}
I_i=I_0\mathrm{exp}\left(\frac{\mathrm{cos}(\theta_i-\theta_0)-1}{\Delta\mu}\right)
.\end{equation}
Here, $\theta_i$ denotes the longitude of probe $i$ in the HEE coordinate system, and $I_i$ is the peak flux from the $i$ probe's measurement. To obtain the maximum signal-to-noise ratio, we took the flux at the peak of the burst. 
Since the frequencies for the four probes are not the same, the peak fluxes from the probes were interpolated to the given frequencies. The peak fluxes at the four viewing sites were fitted for each frequency using Equation \ref{eq:I}. 
From this, we derived the best fits of $I_0$, ${\theta}_0$, and $\Delta\mu$ from the four peak fluxes ($I_i, \theta_i$) at each viewing site, as shown in Figure \ref{obs_flux_pos} (a). Here, ${\theta}_0$ indicates the source longitude that gives the maximal flux, and $\Delta\mu$ represents the shape of the radio emission directivity pattern \citep{2021A&A...656A..34M}. The input errors of the flux are assumed to be 50\% of the peak flux, marked as vertical lines. After applying the density model to convert the frequencies to radial distances, the source positions from the intensity fit are shown as plus signs in Figure \ref{obs_flux_pos} (b).
The longitude uncertainties are represented by the standard deviations of errors derived from the nonlinear least squares fitting of the peak flux curves, considering 50\% uncertainties of the peak fluxes. The source longitudes vary from $-64.1\pm7.3^{\circ}$ at 925 kHz to $-60.7\pm2.8^{\circ}$ at 425 kHz, and the source follows a rather straightforward trajectory.

\subsection{Source positions from the timing method}
\label{sec-obs-timing}

The technique of ToA assumes that delay times are caused by differences in the distances that radio waves travel from the emission region to spacecraft. By using the $\chi^2$ method, the time delays between the spacecrafts and the radio source can be estimated. The source position can then be determined by minimizing the value of $\chi^2$, as shown in Equation \ref{eq:timing}:
\begin{equation}
\label{eq:timing}
\chi^2=\sum\frac{\big(\sqrt{(x_s-x_i)^2+(y_s-y_i)^2}\big/ c+t_0-t^{pk}_i\big)^2}{\Delta t^2_i}
.\end{equation}
Here, $t_0$ is the time of emission and $t^{pk}_i$ represents the peak times of the observations at a given frequency from the spacecraft $i$. The locations of the probes in the ecliptic plane are denoted by $x_i$ and $y_i$. We ignored the latitudes of the four spacecraft since $z_i/r_i$ are small on most days, roughly around $\sim$0.03, 0.01, 0.001, and 0.001 on 05 June 2020 for PSP, SolO, STEREO-A, and WIND, respectively.
While we concentrated on the source's longitude, the radio source's latitude is assumed to be zero, and the source is positioned in the plane of Earth's orbit at ($x_s$, $y_s$, and 0) in the HEE coordinate system. The radio waves are assumed to propagate freely at the speed of light from the source to the probes. Figure \ref{obs_flux_pos} (b) shows the source locations, $x_s$ and $y_s$, determined using the timing method.

The deduced source position is significantly affected by the available time resolution. For instance, a time resolution of $\Delta t=60$ s yields a distance of $c\Delta t=$25.8 R$\odot$. The uncertainties of the positions were determined using a time randomization subset being sampled by the peak times, generating 50 variations where the peak time was varied by $t^{pk}_i\pm \Delta t_i$. Here, $\Delta t_i$ was randomly taken from a normal distribution with a mean of zero and a standard deviation of one. We obtained the average positions ($x_s$, $y_s$, and 0) and the errors from the average standard deviations through nonlinear least squares fitting.

The radio sources at multiple frequencies are located at an average longitude of roughly -60 degrees, similar to the longitude obtained from the intensity fit method. 
However, the source trajectories are not easily distinguishable and they range from ${-47.4}\pm10.3^{\circ}$ to ${-88.6}\pm3.9^{\circ}$, as shown in Figure \ref{obs_flux_pos} (b). The radial distances vary from $23.5\pm1.8$ R$\odot$ at 925 kHz to $46.2\pm2.1$ R$\odot$ at 425 kHz, which are significantly different from the distances deduced from the coronal density model.
Previous observations of interplanetary type III bursts also showed larger heliocentric distances, which increase exponentially with the emission frequencies \citep{1984A&A...141...17B, 1998JGR...103.1923R, 2009SoPh..259..255R}.

\section{Simulations}
\label{sec-sim}

Radio waves that propagate in the turbulent corona and interplanetary space can have their time profiles, source sizes, positions, and directivity altered by the refraction effects of large-scale density gradients and the scattering effects of small-scale density perturbations \citep{1965BAN....18..111F, 1971A&A....10..362S, 1985A&A...150..205S, 2007ApJ...671..894T, 2008ApJ...676.1338T, 2018ApJ...857...82K, 2019ApJ...884..122K}. To investigate the effects of radio wave propagation on source positions, we used a ray-tracing method to simulate radio wave propagation with anisotropic density perturbations, as developed by \cite{2019ApJ...884..122K}.

The simulation treats radio waves as a collection of rays with positions \textbf{r} and wave vectors \textbf{k}. Initially, these rays were considered to originate from a point source in the ecliptic plane, with a given heliocentric angle and distance. The emission frequency can be converted to a heliocentric distance, while the density model $n(r)$ is assumed. Here we applied the density model $n(r)=4.8\times {10}^9 r^{-14} + 3\times {10}^8 r^{-6} + 1.39\times {10}^6 r^{-2.3}$ ($r$ is expressed in solar radii), which is from an analytical approximation (Equation 43 in \cite{2019ApJ...884..122K}) of the Parker density profile \citep{1960ApJ...132..821P}.
As radio waves propagate and undergo scattering in the corona, their positions and wave vectors change, and these can be determined from numerical solutions of the Fokker-Planck equation and Hamilton's equations in an unmagnetized plasma, as described in \cite{2019ApJ...884..122K}. Once fully scattered, the rays arrive at a given sphere beyond which the scattering effects can be considered negligible. The arrival times, final positions, and wave vectors are recorded to produce the time profiles and images. The time profiles of the simulated radio waves after propagation can be presented by the histogram of the rays' arrival times.

The simulated properties of the radio waves mainly depend on four factors: the frequency ratio over the local plasma frequency, the level of density fluctuations $\epsilon$, the anisotropic parameter $\alpha$, and the heliocentric angle $\theta_s$ of the intrinsic source. We considered a fundamental emission frequency of 1.1 times the local plasma frequency. Emissions that are closer to the plasma frequency undergo stronger scattering, resulting in a wider time profile with a longer duration. We note that $\epsilon$ is a relative level of density fluctuation defined as the density fluctuation variance $\langle \delta n^2 \rangle$ normalized by the local density $n$, $\epsilon=\frac{\langle \delta n^2 \rangle}{n^2}=\int S(\mathbf{q}) \frac{d^3 q}{(2\pi)^3}$ \citep{2019ApJ...884..122K, 2023arXiv230805839K}, which depends on the inner and outer scales of the density fluctuations. 
The larger value of $\epsilon$ means a stronger density perturbation and thus stronger scattering, which makes the apparent source size larger and decay time longer.
The anisotropy is defined as the ratio of the parallel and perpendicular wavenumber of the density perturbations, $\alpha=q_{\parallel}/q_{\perp}$. Anisotropic density perturbations with $\alpha=0.2-0.3$, predominantly in the perpendicular direction to the magnetic field, are required to explain observed solar radio bursts \citep{2019ApJ...884..122K, 2020ApJ...898...94K, 2020ApJ...905...43C, 2021A&A...656A..34M}. When $\alpha<1$, radio wave propagation aligns more closely with the radial direction, resulting in a narrower time profile. Stronger anisotropy, corresponding to smaller $\alpha$ values, can reduce the duration of the radio emissions.
With $\epsilon=0.8$, accompanied by the provided inner and outer scales of density fluctuations, and with $\alpha$ set at 0.25, this can explain the time profiles, size, and shift of the radio source at 32 MHz well \citep{2020ApJ...905...43C}. In our simulation, we adopted these specified values of $\epsilon=0.8$ and $\alpha=0.3$, following \cite{2019ApJ...884..122K} and \cite{2021A&A...656A..34M}, and subsequently applied the timing and intensity fit methods to evaluate their effectiveness in the determination of the source positions from the simulated time profiles.


The emitted frequencies were set to span from 425 to 925 kHz in steps of 100 kHz. The corresponding emission region is situated at a heliocentric distance ranging from 16.4 to 8.6$R\odot$. In our simulations, we traced $1 \times 10^6$ photons through the corona until all rays arrived at a sphere with a radius of 1 AU. Initially, the rays were located at a heliocentric angle of ${-50}^{\circ}$. The resulting apparent source is located at $\sim{-60}^{\circ}$, which agrees with the average longitude of the radio source inferred from observations of the interplanetary type III burst.
In order to match the positions of four probes, we collected rays at a 1 AU sphere with viewing angles of -149, 42, 0, and -71 degrees for PSP (P1), SOLO (P2), STEREO-A (P3), and WIND (P4), respectively. All probes are nearly lying in the ecliptic plane, and their latitudes are ideally close to zero. To obtain a better time profile with less statistical errors, we set the latitude of the collection positions to 0.85<$\cos\phi$<1 and centered the longitudes at viewing angles with a spread of 10 degrees.
The rays that arrived at 1~AU with a wave vector ($k_x$, $k_y$, $k_z$) and position ($r_x$, $r_y$, $r_z$) were traced back to the locations of the probes. The number of photons arriving at each viewing site varied over time, as shown in Figure \ref{rays_flux_pos} (a).

We fit the peak fluxes at different viewing longitudes using Equation \ref{eq:I} (Figure \ref{rays_flux_pos} (b)). The uncertainty was set to 50\% of the flux, and a statistical Poisson weighting was applied to the intensity fit. The longitude ($\theta_0$) at which the flux reached its maximum is regarded as the most probable direction of the source position. 
The heliocentric radial distance was assumed to be the same as the initial radial distance deduced from the density model for each frequency. Source positions ($r^{\rm{I-fit}}$, $\theta^{\rm{I-fit}}_0$) estimated from the intensity fit method are shown in Figure \ref{rays_flux_pos} (c). 
We also used the timing method to determine the source position. The times at which the peak flux occurred were identified and fitted using Equation \ref{eq:timing}. The source positions determined from timing are marked as triangle symbols in Figure \ref{rays_flux_pos} (c).

The apparent source position was inferred from the image centroid and the direction of the centroid. The centroid direction corresponds to the emission directivity peak, defined as $\mu=k_z/|\mathbf{k}|$, where $\mathbf{k}$ is the wave vector and the $z$ direction is the Sun-Earth direction.

\begin{figure}[h!]
\centering 
\includegraphics[width=9cm]{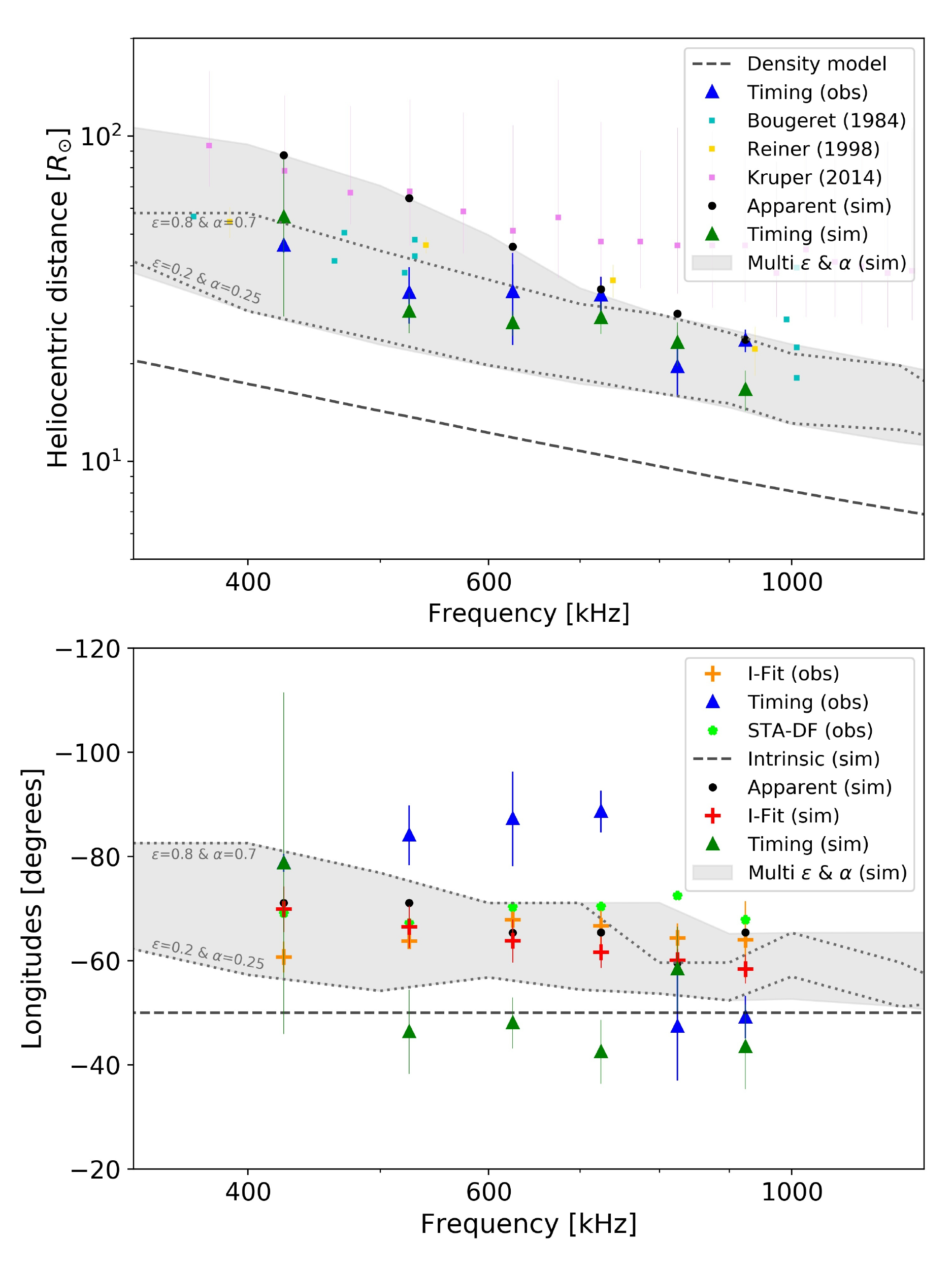}
\caption{Heliocentric distances (upper panel) and longitudes (lower panel) of the source were deduced from observations of four spacecraft (obs) and simulations (sim) of radio wave propagation for anisotropic scattering effects. The dashed line represents the radial distance calculated from the density model, which corresponds to the heliocentric distances of the true source. The referenced distances of the IP type III bursts from spacecraft observations were obtained from previous studies \citep{1984A&A...141...17B, 1998JGR...103.1923R, 2014SoPh..289.4633K}. The source positions deduced from the intensity fit and timing method, based on the intensity profiles in the simulations (with $\epsilon$=0.8 and $\alpha$=0.3), are depicted as red plus symbols and dark green triangle symbols, respectively. The gray shadow indicates the apparent source positions obtained from simulations considering a range of $\epsilon$ values from 0.2 to 0.8 and $\alpha$ values from 0.25 to 0.7. The green symbols (STA-DF) indicate the directions of this type III burst as determined by the  DF measurement of STEREO-A.
\label{freq_r_lon}}
\end{figure}

\section{Discussion}
\label{sec-sum}

To determine the source positions of interplanetary type III bursts, there are various methods that can be implemented. One such method is triangulation using DF analysis, which requires measurements from at least two spacecraft. Another method is the timing method, which can be used when there are three or more spacecraft measurements available. In addition, it is possible to deduce the direction of the radio source from a single spacecraft DF measurement, or from the intensity fit that relies on the absolute flux from at least three spacecraft measurements. 
For this study, we have utilized both the intensity fit and timing method to determine the source positions of an interplanetary type III burst. This is the first time these methods have been applied to radio wave simulations with anisotropic scattering effects and compared with the measurement results.
The true source position remains elusive based on observations. However, in scattering simulations, an intrinsic source position is given, and apparent source positions can be inferred after the radio wave undergoes scattering. While observations lack the necessary information to fully evaluate the validity of those methods, simulations grant us a clear understanding of the source's positions before and after their propagation and facilitate the comparison of these methods.

Four space-based radio instruments, namely PSP/RFS, SolO/RPW, STEREO-A/WAVES, and WIND/WAVES, provided the four viewpoints from which the burst was detected. By exploiting the significant separation between these spacecraft, the source directions were determined through intensity fits performed on peak intensities from multiple viewing locations.
The timing method relies on the arrival time difference between two spacecraft to locate the source position, assuming that the radio waves propagate in straight lines. Timing measurements are ideal for determining source locations when the spacecraft and radio source are at vastly different distances from each other, as the propagation times can exceed time resolutions.
One notable limitation is the impact of time resolution on deducing the source position using the timing method. To address this, we considered the time resolutions of four spacecraft and provide uncertainties using time randomization subset sampling. Another limitation is that the most intense part of the burst may not be observed, which is characterized by its peak intensity on the dynamic spectrum. We considered the intensity uncertainty and derived standard deviations of errors from nonlinear least squares fitting for the intensity fit method.
During this event, the latitudes of the four spacecraft were nearly situated around 0 degrees, whereas the source's latitudes inferred from STEREO DF measurements varied within 10 degrees out of the ecliptic plane. Given these circumstances, it is reasonable to perform an intensity fit, which is based on the assumption that the source’s latitude is zero, as demonstrated by \cite{2021A&A...656A..34M}, and timing to determine the source's position projected in the ecliptic plane.

The heliocentric distances and directions in the ecliptic plane at each frequency are shown in Figure \ref{freq_r_lon}. 
The radio source's longitudes deduced from the intensity fit are comparable to the longitudes from the DF measurement (vary within 10 degrees) and the radio sources follow a rather straightforward trajectory. 
From timing, the radio source is found to be located further away with a decreasing frequency but it did not follow a specific direction. The heliocentric distances follow a power law function with the frequency of $r=(20.66\pm2.57)\times f^{-0.91\pm0.20}$. Our measurement is consistent with previous studies, which have also found heliocentric distances estimated from radio triangulation to be larger than the ones computed from coronal density models \citep{1961ApJ...133..983N, 1977SoPh...55..121S} \cite[e.g., from][]{1984A&A...140...39S, 1984A&A...141...17B, 1998SoPh..183..165L,  2009SoPh..259..255R, 2014SoPh..289.4633K, 2022ApJ...938...95B}.

The propagation of radio waves in interplanetary space is influenced by scattering on density fluctuations, leading to changes in time profiles, directivity, and source positions. To simulate radio-wave propagation, we used the ray-tracing method and predicted intensity profiles at various viewing angles. The sequential arrival of rays and the relative peak intensity at multiple viewing angles show similarities to observations. 
In Figure \ref{rays_flux_pos} (c), we compare the source positions from imaging with those deduced from the intensity fit and timing method, finding that the direction deduced from the intensity fit is close to that of the apparent source, which deviates from an angle from the given intrinsic source and seems to be closely aligned with the Parker spiral magnetic field. However, the heliocentric distances and longitudes determined from the timing method do not match the apparent sources, suggesting that the source positions may be underestimated.

The time delays between two observers depend on two factors: the propagation distance from the source to the observer and the scattering time that varies as the rays travel toward distinct viewing sites. Our simulations suggest that anisotropic scattering leads to larger heliocentric distances for the apparent position than expected from the intrinsic radio source. While the radial distances of the radio source at 525 kHz from imaging and being deduced from the timing method show 64.6 and 35.4 $R_{\odot}$, the initial radial distance is 13.7 $R_{\odot}$ from the coronal density model. 
The observed source positions have been displaced away from the locations where emission is generated. The radial distances deduced from observations exceed the values predicted by density models, indicating alignment with radio-wave propagation affected by anisotropic scattering, which could lead to an apparent position at a heliocentric distance farther from the Sun.

The active region 12765 is responsible for initiating this type III burst, and the PSP spacecraft is located behind it, as introduced from \cite{2022A&A...657A..21S}. They also suggest that PSP can detect the radio emission from behind due to the radio producing electrons propagating in a dense loop.
We note that the radio emissions can be detected backward owing to refraction and scattering of radio waves, and the directivity distribution reveals that these emissions encompass wide viewing angles and are centered approximately -60 degrees apart from the Sun-Earth directions.

Correcting the source position from the radio-wave propagation with anisotropic scattering effects is challenging. Attempts have been made to include the effects of refraction, such as \citet{2010ApJ...720.1395T}, but we find it difficult to diagnose the anisotropy parameter and the relative density fluctuation level of the interplanetary turbulence from only the dynamic spectra, and scattering is event-dependent.

Within this investigation, we inferred and compared the longitudinal position of an interplanetary type III burst's source by using DF measurements from STEREO, the intensity fit, and timing method. Our primary scope did not encompass source positions situated out of the ecliptic plane. 
We have also ignored the evolution of the radio source itself and instead directed our attention toward the propagation of radio waves subsequent to their emission. The inhomogeneity can affect the electron beam dynamics, plasma waves, and consequently the generation of radio waves. Our primary focus centers on understanding the effects of radio wave propagation -- how emitted radio waves propagate from the source to observer in the nonuniform interplanetary space.

Many factors may affect the estimation of source positions, such as time resolutions, intensity uncertainties, frequency differences, assumptions about the radio source location, fundamental or harmonic emissions, and the intrinsic source size. We simplified the simulation by assuming a point source and not considering the actual source size, while the observed intensity profile is the result of the convolution between the intrinsic emission and broadening due to scattering \citep{2020ApJ...905...43C}. Additionally, each spacecraft may detect a different section of the extended source, which may have a significant size or comprise multiple emitting regions. 
Another important factor that affects the determination of the source position is whether the radio waves experience full scattering effects during their propagation through all spacecraft. The proximity of PSP to the radio source does not guarantee full scattering effects during radio-wave propagation, and some scattering effects may still occur after passing through the spacecraft.

\begin{acknowledgements}
      This work is supported by STFC consolidated grant ST/T000422/1. NC acknowledges funding support from CNES and from the Initiative Physique des Infinis (IPI), a research training program of the Idex SUPER at Sorbonne Universit\'{e}. XC thanks NSFC Grant 12003048.
      The authors would like to thank the PSP/RFS, SolO/RPW, STEREO/WAVES, and Wind/WAVES teams for making the data available.
      Solar Orbiter is a space mission of international collaboration between ESA and NASA, operated by ESA.
      The FIELDS experiment on the Parker Solar Probe spacecraft was designed and developed under NASA contract NNN06AA01C.
\end{acknowledgements}

\bibliographystyle{aa}
\bibliography{pos_IP3}
\end{document}